\documentclass[fleqn,twoside]{article}
\usepackage{espcrc2}
\usepackage{graphicx}
\newcommand{\AmS}{{\protect\the\textfont2
  A\kern-.1667em\lower.5ex\hbox{M}\kern-.125emS}}

\title{The Anatomy of the Knee and Gamma-Families}
\author{A.D.Erlykin\address{P.N.Lebedev Physical Institute  
        Leninsky prosp. 53, Moscow 119991, Russia}
        \thanks{E-mail: erlykin@sci.lebedev.ru},
        A.W.Wolfendale\address{Physics Department, University of Durham,
        South Road, Durham DH1 3LE, UK}}

\begin{document}

\begin{abstract}
It is shown that the fine stucture of the cosmic ray energy spectrum in the knee 
region, if explained by the Single Source Model (~SSM~), can, in principle, be clearly 
revealed and magnified in the size spectrum of extensive air showers (~EAS~) associated
 with gamma families. Existing experimental data on EAS at mountain level give support 
to this hypothesis.     
\vspace{1pc}
\end{abstract}

\maketitle

\section{Introduction}
The combination of the original EAS technique with photosensitive materials for the 
study of the microstructure of EAS cores was proposed long ago 
\cite{Brik,Rapp,Grig}. Since then, in spite of both methods being well 
developed and widely used in cosmic ray (~CR~) studies they have remained basically 
independent of each other. However, there have been several attempts to combine them 
\cite{Smor,Mata,Dak1,Dak2,Ohta,Shau1,Adam,Kawa} and these combinations have brought 
interesting results. Inspired by these results a few new projects have been proposed 
to use large stacks of X-ray films within EAS arrays at mountain altitudes, where 
the intensity and the energy content of both EAS and gamma-hadron families is much 
higher than at sea level \cite{Aoki,Saav,Amur}. In this paper we draw an attention to
a promising opportunity offered by this technique.       

\section{The Single Source Model of the Knee}

The Single Source Model was proposed by us to explain the fine structure of EAS size
 spectra in the knee region \cite{EW1,EW2}. Besides the sharp knee at EAS sizes $N_e$
corresponding to a primary energy of about 3-4 PeV, a second sharp 'peak' 
(~in $log(IN_e^3)$ vs $logN_e$ coordinates~) has been found at $N_e$ corresponding to 
an energy in the range 12-16 PeV. We attribute these structures: the knee and the 
second 'peak', to a sharp cutoff of the rigidity spectrum for primary CR nuclei 
accelerated by a single, nearby and recent supernova. Initially, we associated the knee
 with the cutoff of oxygen nuclei and the second 'peak' - with iron nuclei, although 
we cannot exclude now that they
are caused by the energy cutoffs of helium and oxygen \cite{EW2,EW3,EW4}. Due to the 
flat spectrum of the Single Source (~$\gamma = 2$~) and the sharp cutoffs the peaks
in fact look like lines in the EAS size spectra. However, the 'intensities' of these 
lines are low and they are hardly seen above the background. We look for the 
possibility of increasing the 'signal to noise ratio' in the ground-based observations
using EAS associated with gamma families.

\section{Gamma families initiated by primary nuclei}

The standard attitude is that if one selects EAS above a threshold size $N_e^{thr}$ or 
a gamma family with a threshold total energy of gamma quanta $\Sigma_i^n E_i$ then 
they are induced preferentially by primary protons. It is true, but cannot be 
extended to shower sizes $N_e$ much higher than the threshold $N_e^{thr}$. In Figure 1
we show the probability for primary protons and iron nuclei of different energies 
creating EAS within a zenith angle interval $\Theta < 30^\circ$ containing gamma quanta
 and gamma families at the Tien-Shan altitude (~3340 m a.s.l.~). Calculations have been
 made using the CORSIKA 6.014 package with the QGSGET interaction model \cite{Heck}. 
The multiplicity $n_\gamma$ of gamma-quanta in the family has been taken as 
$n_\gamma \geq 1$ and $n_\gamma \geq 2$; the minimum energy of gamma-quanta
was 4 TeV. It is seen that: \\
(i) in the wide energy region the energy dependence of the probability $P(E)$ is 
stronger then the direct proportionality $P \propto E$;\\
(ii) at energies above 10 PeV, where the probability for protons to produce the 
gamma family approaches 1, the corresponding probability for nuclei also approaches 1.
 Therefore, beyond the knee, where according to many experiments the primary CR mass 
composition is enriched by heavy nuclei, they will be also efficient in producing 
gamma-families.
\begin{figure}[htb]
\begin{center}
\includegraphics[height=7.5cm,width=7.5cm,angle=-90]{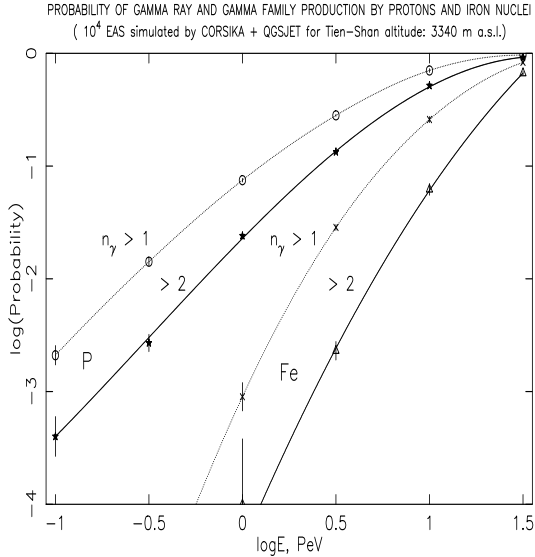}
\caption{\footnotesize The probability for protons (P) and iron nuclei (Fe) producing 
gamma-families with multiplicity $n_\gamma \geq 1$ and $n_\gamma \geq 2$ at 
Tien-Shan altitude, 3340 m a.s.l. Open circles: P,$n_\gamma \geq 1$, stars - P,
$n_\gamma \geq 2$, crosses - Fe,$n_\gamma \geq 1$, triangles - Fe,$n_\gamma \geq 2$. 
Dashed and full lines are polynomial fits of the data. The threshold energy of gamma 
quanta is 4 TeV.}
\end{center}
\label{fig:fam1}
\end{figure}    

\begin{figure}[h]
\begin{center}
\includegraphics[height=7.5cm,width=7.5cm]{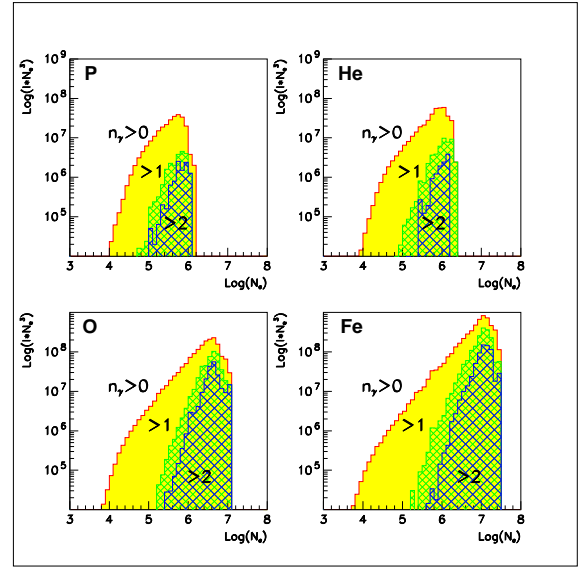}
\caption{\footnotesize Size spectra of EAS for P, He, O and Fe nuclei (~indicated 
inside the graphs~), invoked in the SSM. The histogram with the highest statistics
(~$n_\gamma \geq 0$~) are for all EAS, histograms with intermediate statistics - EAS 
containing gamma quanta, i.e. showers with $n_\gamma \geq 1$, histograms with lowest 
statistics - EAS containing gamma families, i.e. showers with $n_\gamma \geq 2$.}  
\end{center}
\label{fig:fam2}
\end{figure}    
\section{Size spectra of EAS containing gamma families}

The SSM in common with many other models implies that different CR nuclei are 
accelerated by a single source up to the same maximum rigidity $R_{max}$. In 
particular, in the SSM with helium dominating at the knee $logR_{max}, GV = 6.2$. 
In Figure 2 we show the size spectra of EAS produced by protons
$P$, helium $He$, oxygen $O$ and iron $Fe$ nuclei, which can be accelerated by our
Single Source. The minimum rigidity $R_{min}$ has been taken as 0.1 PV. The input 
conditions (~CORSIKA6.014, QGSJET, $\Theta \leq 30^\circ$, $E_\gamma \geq 4TeV$ ) were 
the same as described in the previous section. Energy spectra within the rigidity 
intervals have been taken as $\sim E^{-2}$. Showers are shown both containing high 
energy gamma quanta and families, i.e. for $n_\gamma \geq 1,2$, and without them.   
It is seen that due to the strong energy dependence of the probability $P(E)$ the size 
spectra of EAS containing gamma quanta and gamma families are flatter than those for 
all showers. Since the EAS size cutoff is sharp, these spectra shown in our coordinates
 $log(IN_e^3)$ vs. $logN_e$ look even more like spectral lines, than the spectra of 
all showers.

\section{The size spectrum of EAS from the Single Source}

In Figure 3 we show the spectrum of all showers and showers containing gamma quanta and
 gamma families for our SSM, i.e. with energy independent abundance $\Delta_A$ of 
different nuclei as $\Delta_P = 0.004$,  $\Delta_{He} = 0.711$,  $\Delta_{O} = 0.210$ 
and $\Delta_{Fe} = 0.075$ for fixed rigidity.  
\begin{figure}[htb]
\begin{center}
\includegraphics[height=7.5cm,width=7.5cm]{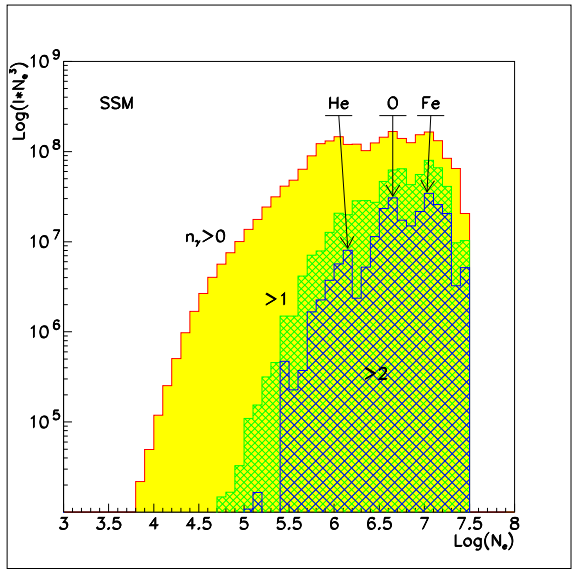}
\caption{\footnotesize The size spectrum of EAS from the Single Source, which emits P, 
He, O and Fe nuclei with constant abundances: 0.004 for P, 0.711 for He, 0.210 for 
O and 0.075 for Fe. The relevant histograms for EAS are the same as in Figure 2. The 
location of the peaks is indicated.}
\end{center}
\label{fig:fam3}
\end{figure}    
Due to sharper peaks in the individual spectra related to the energy cutoffs of 
different nuclei in CR from the Single Source these peaks are more distinct in the 
spectrum of EAS containing gamma families rather than in the total spectrum of all EAS.
As can be seen in Figure 3 the excess over the smooth background in the spectrum of 
all showers in the logarithmic scale is about $\Delta(logIN_e^3) = 0.18\pm 0.01$ 
(~signal/noise ratio $S/N\simeq 0.5$~), while that in the spectrum of showers 
containing gamma families is 
about $0.36\pm 0.02$ (~$S/N\simeq 1.3$~). Of course, the statistics of the showers with
 families is less than that of all showers and, besides the smooth background of the 
Single Source, there is a background from other sources, which reduces the $S/N$ 
ratio. However, the estimates based on our simulations show that the confidence level
of detecting the signal in the size spectrum of EAS with families is about the 
same as in the spectrum of all EAS. 

There is experimental evidence that in the knee region the size spectrum of EAS with
 gamma families observed at mountain altitude has sharp peaks, which are more 
clearly visible than irregularities in the spectrum of all showers \cite{Shau2}. Since 
this Tien-Shan experiment has the highest statistics of such showers in the world, 
this result deserves further analysis and the technique of the combined use of
EAS detectors with X-ray films - a further development and longer exposure.   

Reliable distinction of proton- and helium- induced peaks can help to make a choice 
between models of the cosmic ray origin in which cut-off energies in the spectra of 
constituent nuclei are proportional to their charges \cite{Voelk} or masses \cite{DDD}.

\section{Conclusion}

It is shown that if there is a sharp cutoff in the rigidity spectrum of accelerated 
nuclei, as in the Single Source Model of the knee, then the size spectrum of the EAS
containing gamma families can have more distinct peaks corresponding to the cutoffs
of different nuclei than the spectrum of all EAS. There is experimental evidence
for the existence of such peaks and the present paper is to some extent inspired by 
this unique observation.

\end{document}